\documentclass[twocolumn,notitlepage,prl,superscriptaddress]{revtex4-1}
\usepackage{amsmath}
\usepackage{amsfonts}
\usepackage{amssymb}
\usepackage{graphicx}
\usepackage{sidecap}
\usepackage{braket}

\usepackage{blindtext}
\usepackage{subfigure}
\usepackage{lipsum}
\usepackage{fullpage}

\usepackage{appendix}
\usepackage{subfigure}
\usepackage{color}
\usepackage[dvipsnames]{xcolor}
\usepackage{physics}

 \usepackage{amsthm}

\begin{document}
\title{Non-Markovian super-superradiance in a linear chain of up to 100 qubits}

\author{Fatih Dinc}
\thanks{{Present adress: Department of Applied Physics, Stanford University, Stanford, CA, USA \\ Email: fdinc@stanford.edu}}
\affiliation{Perimeter Institute for Theoretical Physics, Waterloo, Ontario, N2L 2Y5, Canada}
\author{Agata M. Bra\'nczyk}
\affiliation{Perimeter Institute for Theoretical Physics, Waterloo, Ontario, N2L 2Y5, Canada}

\begin{abstract}
We study non-Markovian enhancement effects in the spontaneous emission of a collective excitation in a linear chain of up to 100 qubits coupled to a 1D waveguide. We find that for a critical separation of qubits, the system exhibits super-superradiant (SSR) behavior leading to collective decay stronger than the usual Dicke superradiance. Here,  time-delayed coherent quantum feedback effects are at play on top of the usual Dicke superradiance effects. We find a linear scaling for the SSR decay rate with increasing qubit number $N$ such that $\Gamma_{\rm SSR} \sim 2.277 N \gamma_0$, where $\gamma_0$ is the single emitter decay rate to a one-dimensional waveguide, as opposed to $\Gamma_{\rm Dicke}\sim N \gamma_0$ for Dicke superradiance. The SSR decay rate can be tuned with qubit separation distance and may therefore have application for quantum technologies.
\end{abstract}

\maketitle

Surprising phenomena can emerge when multiple quantum systems coordinate their behaviour. A well-known example of this---Dicke superradiance \cite{andreev1980collective,dicke1954coherence}---occurs when a group of $N$ quantum emitters, excited in a symmetric state, exhibits  \emph{enhanced} spontaneous emission \cite{fleury2013enhanced,john1995localization,vats1998non}. This phenomenon, predicted when the distance between emitters is small compared to the wavelength of radiation,  can be understood as \emph{collective} behaviour of the entire system. Curiously, enhanced emission  can also occur  when the emitters are macroscopically separated  \cite{zheng2013persistent,sinha2019non,dinc2019exact,zhou2017single}; a regime in which collective behaviour might not be expected. What's more, the emission can become  more enhanced than Dicke superradiance; earning it the name  \emph{super-superradiance} (SSR) \cite{dinc2019exact}. This highly counter-intuitive phenomenon can be explained by an effect that has recently gained attention in other contexts \cite{grimsmo2015time,guimond2017delayed,pichler2016photonic,calajo2019exciting,pichler2017universal,whalen2017open,chalabi2018interaction}:  \emph{time-delayed coherent quantum feedback}. In this paper, we study the effects of time-delayed coherent quantum feedback on the \emph{super-superradiant} collective decay rate of a linear chain of qubits coupled to a one-dimensional (1D) waveguide.

The 1D geometry of the waveguide enhances scattering and is thus  particularly favourable to study time-delayed coherent quantum feedback. While linear chains of qubits in 1D waveguides have been studied extensively, both theoretically  \cite{song2018photon,das2018photon,liao2015single,tsoi2008quantum,albrecht2019subradiant,zhou2017single} and experimentally \cite{goban2015superradiance,hood2016atom,goban2014atom}, the literature mostly focuses on regimes where the Markovian approximation is valid, i.e. time retardation effects are negligible. Time-delayed coherent quantum feedback, however, occurs when qubits are separated by distances in which the Markovian limit is no longer valid. One must, therefore,  take into account non-Markovian dynamics of qubit-light interactions. This was done in waveguide QED for a single qubit in \cite{calajo2019exciting,tufarelli2014non,ramos2016non,fang2018non} for two qubits in \cite{zheng2013persistent,sinha2019non} and three qubits in \cite{dinc2019exact}.  Here, we consider up to 100 qubits.

\begin{figure}
    \centering
    \includegraphics[width=\columnwidth]{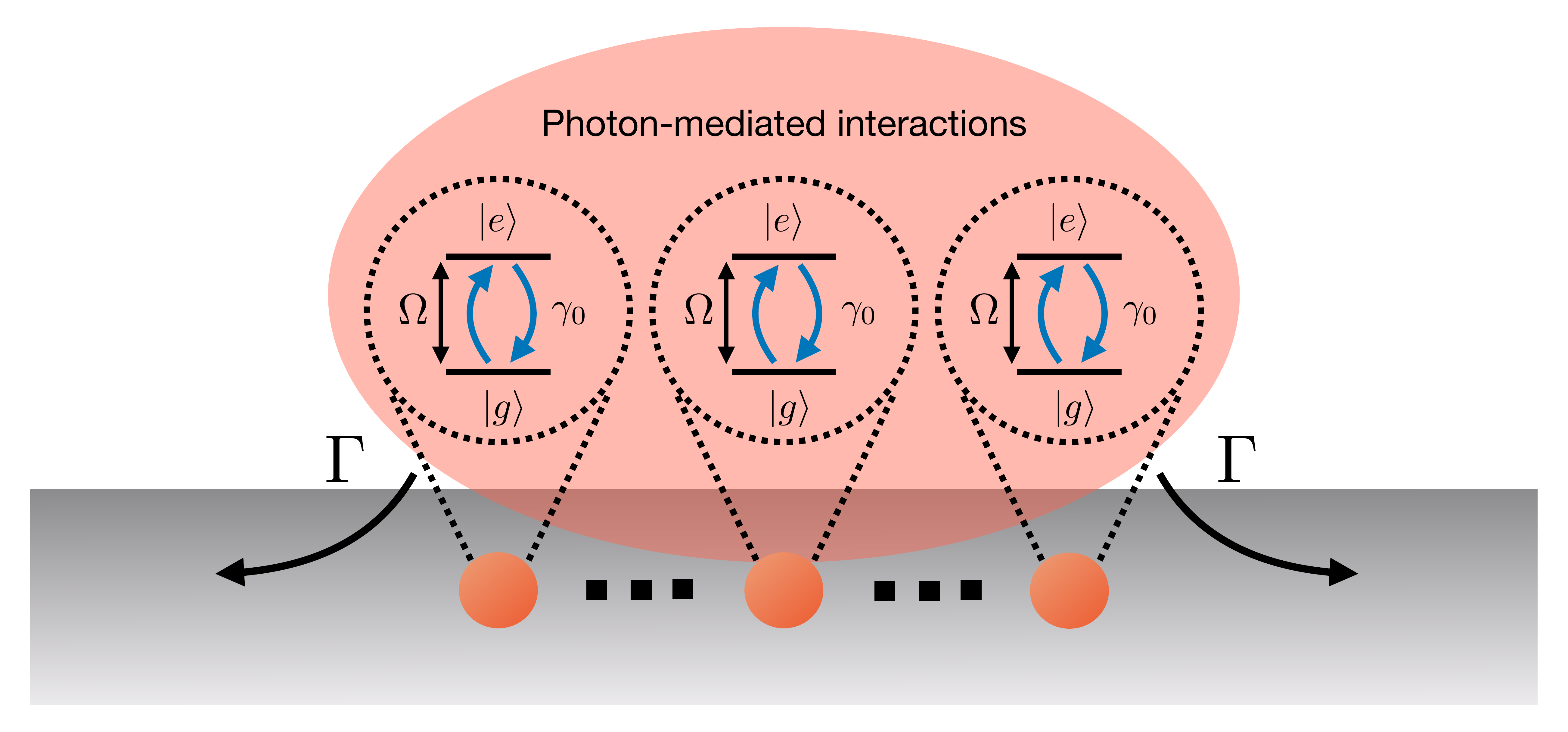}
    \caption{The linear chain of $N$ qubits coupled to a 1D waveguide. The photon-mediated interactions between qubits lead to a collective behavior of the system described by collective decay rates. An initially excited system decays through these decay modes.}
    \label{fig:fig1}
\end{figure}

We consider a linear chain of $N$ identical qubits separated by a distance $L$ and coupled to a 1D waveguide as shown in Fig. \ref{fig:fig1}. We study SSR by computing the collective decay rates of the $N$-qubit system. Our method of choice is the real-space approach \cite{dinc2019exact}. In this approach, one first identifies the transfer matrix for the system, then uses that matrix to write down and solve a characteristic equation for the collective decay rates. The real-space approach has the following strengths for this problem: (1) No need to model time-dynamics, as the collective decay rates can be found from steady-state single-frequency solutions (energy eigenstates for the whole system); (2) No need to identify the multi-qubit superposition states that corresponds to SSR (in fact, one does not need to consider the qubit excitation subspace at all); and (3) Energy eigenstates can be found via a recursive transfer matrix method. These strengths of the real-space approach make the problem significantly more tractable for large-$N$ systems.

We start by identifying the transfer matrix. For a plane wave with frequency $k$, incident from far left, the transformation  that relates the transmitted and reflected photon coefficients, $t$ and $r$ respectively, to the incident photon is 
\begin{align} \label{eq:transfermatrix}
    \begin{pmatrix}
    1 \\
    r
    \end{pmatrix}
    = T^N
    \begin{pmatrix}
    t \\
    0
    \end{pmatrix}.
\end{align}
Here, $T$ is the transfer matrix corresponding to a unit cell, which consists of inter-qubit propagation and single qubit-light interaction. {Following \cite{dinc2019exact}}, $T$ can be written as
\begin{align}
    T=
    \begin{pmatrix}
    1 + i\frac{\gamma_0}{2\Delta_k} & i \frac{\gamma_0}{2\Delta_k} \\
    - i \frac{\gamma_0}{2\Delta_k} &  1 - i\frac{\gamma_0}{2\Delta_k}
    \end{pmatrix}
    \begin{pmatrix}
    e^{-ikL} & 0 \\
    0 & e^{ikL}
    \end{pmatrix},
\end{align}
where $L$ is the distance between two adjacent qubits, $\Delta_k = E_k - \Omega$ is the detuning between the photon energy $E_k$ and the qubit energy separation $\Omega$,  and $\gamma_0$ is the decay rate of a single qubit to the 1D waveguide. When deriving this transfer matrix, we use the real-space Hamiltonian \cite{dinc2019exact,tsoi2008quantum,shen2005coherent}, where the light-matter interactions take the form of a delta-function at the qubit position. For the scope of this paper, we assume $\gamma_0/\Omega \ll 1$ in alignment with the rotating-wave approximation performed at the Hamiltonian level. For the remainder of this paper, we set $\hbar=1$ and $v_g=1$, where $v_g$ is the group velocity of the photons inside the waveguide. 

We now determine the characteristic equation for the collective decay rates by setting the first element of the transfer matrix in (\ref{eq:transfermatrix}) to zero:  $(T^N)_{11}=0$. The decay rates, $\Gamma$, can then be obtained via a complex rotation followed by a scaling from the poles $\Delta_k^{(p)}$ of the scattering parameters, i.e., $\Gamma = 2i \Delta_k^{(p)}$ \cite{dinc2019exact} (the real part of $\Gamma$ corresponds to the physical decay rate and the imaginary part {is responsible for} a characteristic frequency shift). The factor of $2$ comes from the fact that we are interested in population decay rates. With this definition, the characteristic equation of the poles for an $N$ qubit system is
\begin{equation}
    f(\Delta_k) = \Delta_k^N (T^N)_{11},
\end{equation}
where the multiplication with $\Delta_k^N$ results in a characteristic equation that is polynomial in $\Delta_k$ and $e^{ikL}$ (hence analytic in the complex plane). As a result, the problem of finding the collective decay rates comes down to finding the roots of the function $f(\Delta_k)$. Solving this characteristic equation for a single qubit and performing the rotation, we find the single emitter decay rate to be $\gamma_0$ as expected.

One can consider this problem of finding the collective decay rates in two different regimes: the Markovian regime in which the qubits are microscopically separated ($L \sim O(\Omega^{-1})$) and the propagation time of the photons within the system is negligible; and the non-Markovian regime in which the qubits are macroscopically separated ($L \sim O(\gamma_0^{-1})$) and the propagation time of the photons within the system is not negligible. The superradiance condition in both regimes is $\Omega L = \pi n$, where $n$ is an integer \cite{zhou2017single,dinc2019exact}. 

In the Markovian regime,  the propagation phase acquired by the photon between two qubits can be linearized, $kL \simeq \Omega L$  \cite{dinc2019exact,zheng2013persistent}. In this regime, $f(\Delta_k)$ can be approximated by an $N$th degree polynomial in $\Delta_k$, since the exponential terms in the transfer matrix can be replaced by a constant $e^{ikL}\simeq e^{i\Omega L}$. The $N$ zeros, which  correspond to $N$ collective decay rates,  can be found analytically. When the superradiant condition is satisfied, the superradiant decay rate is $N\gamma_0$ and all other decay rates are zero \cite{dinc2019exact}. This is the case of Dicke superradiance.

In the non-Markovian regime, the photon's propagation phase can no longer be linearized, so the Markovian approximation is no longer valid. In this regime, $f(\Delta_k)$ can no longer be approximated by a polynomial (it depends on $e^{ikL}=e^{i\Omega L (1+\Delta_k / \Omega)}$ as well). Relaxing the linearization condition leads to time-retardation effects inside the multi-qubit system. A discussion on how the linearization $kL\simeq \Omega L$ is linked to the Markovianity assumption can be found in \cite{dinc2019exact} (in particular, in Sections VI and VIII). Consequently, $f(\Delta_k)$ has  infinitely many zeros that correspond to infinitely many collective decay rates, which can be divided into two categories: those that tend to Markovian decay rates in the limit $L\to 0$ (we call these Markovian-like) and those that tend to infinity in the same limit (we call these exclusively non-Markovian). 

SSR refers to the phenomenon that a Markovian-like decay rate surpasses the Dicke superradiance decay rate $\Gamma_{\rm Dicke}= N\gamma_0$. To achieve maximum SSR, the qubits must be separated by a critical distance $L_c$ (as pointed out in \cite{sinha2019non}) as well as satisfy the condition $\Omega L_c = n\pi$. For any $L$ and  $\Omega L = n\pi$, $f(\Delta_k)$ has $N-1$ zeros that correspond to $\Gamma=0$ (this occurs when $\Delta_k=0$). This leaves only one Markovian-like decay rate which could potentially become SSR, i.e. the non-zero $\Gamma$ closest to the origin.

In our analysis, we therefore first assume that $\Omega L = n \pi$ and set $e^{ikL} =e^{i\Omega L (1+\Delta_k / \Omega)} = (-1)^n e^{i \Delta_k L}$. We then use a numerical root finding algorithm to find the first $N$ zeros of $f(\Delta_k)$ that are closest to the origin. $N-1$ of such zeros are found to be  within the vicinity of $\Delta_k =0$,  deviating only by small amounts due to numerical imprecision. The $N$th zero, labelled as $p$, gives the  superradiant decay rate via the Wick-like rotation $\Gamma_u = 2i p$. The real part of $\Gamma_u$ gives the physical decay rate. We define the  SSR decay rate as $\Gamma_{\rm SSR} = \max_{L} \text{Re}[\Gamma_u]$, and perform the maximization numerically. As a sanity check, we then plot $\log(|f(\Delta_k)|)$  for $\Delta_k$ near $\Gamma_{\rm SSR}$ and compare it with the values found numerically.

\begin{figure}[b!]
    \centering
    \includegraphics[width=\columnwidth]{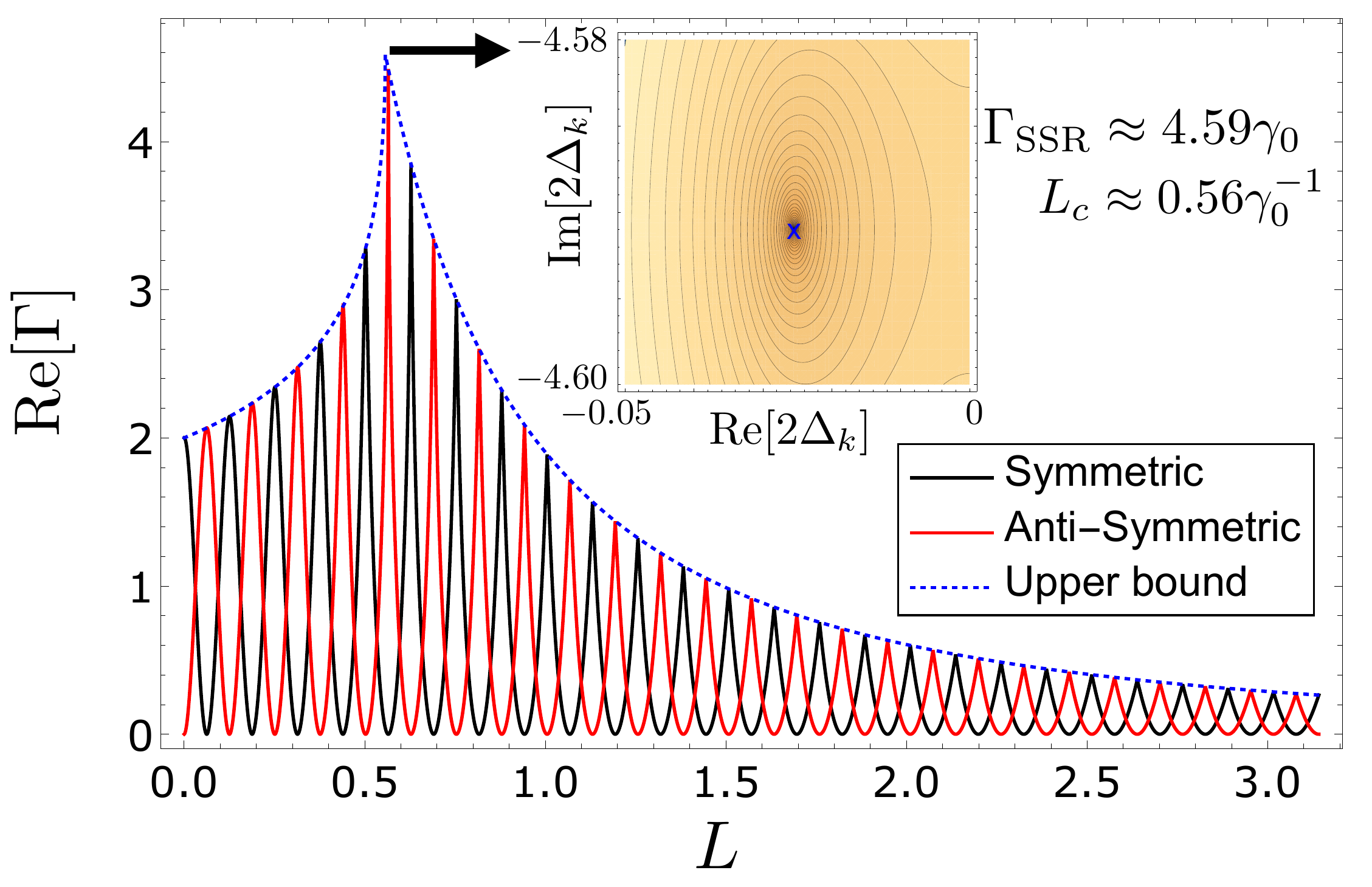}
    \caption{The physical decay rate, given by the real part of $\Gamma$ (in units of $\gamma_0$) for $N=2$ qubits in the non-Markovian regime as a function of qubit separation $L$ (in units of $\gamma_0^{-1}$). {Black ($\text{Re}[\Gamma]=2\gamma_0$ at $L=0$) and red lines ($\text{Re}[\Gamma]=0$ at $L=0$)} correspond to symmetric and anti-symmetric decay rates for $\Omega=50\gamma_0$. The {dashed blue} line gives the  upper bound obtained in the limit $\Omega \to \infty$ of the superradiant decay values. All numerical values are normalized w.r.t. $\gamma_0$. Inset: $\log(|f(\Delta_k)|)$ for $L=L_c$ for $\Delta_k$ close to the SSR pole.}
    \label{fig:fig2}
\end{figure}

Fig. \ref{fig:fig2} shows the   symmetric and anti-symmetric  decay rates (i.e. those that couple to symmetric and anti-symmetric superposition states) for $N=2$ qubits in the non-Markovian regime as a function of qubit separation $L$ for finite $\Omega$, enveloped by an upper bound given by $\Omega\rightarrow\infty$ (for $L=0$, the symmetric decay rate couples to the Dicke state). The symmetric and anti-symmetric decay rates follow the same trend as in Fig. 4(d) of \cite{zheng2013persistent}. The numerical values $\Gamma_{\rm SSR}\approx4.59\gamma_0$ and $L_c \approx  0.56 \gamma_0^{-1}$ match perfectly with the findings of \cite{sinha2019non}. When the condition $\Omega L_c = n\pi$ is satisfied, the peak of the curve for the symmetric decay rate lines up with the peak of the {dashed blue} envelope function, as is the case in Fig. \ref{fig:fig2}. If the  condition is not satisfied, the system acquires a maximum decay for a distance closest to $L_c$. 

\begin{figure}[t!]
    \centering
    \includegraphics[width=\columnwidth]{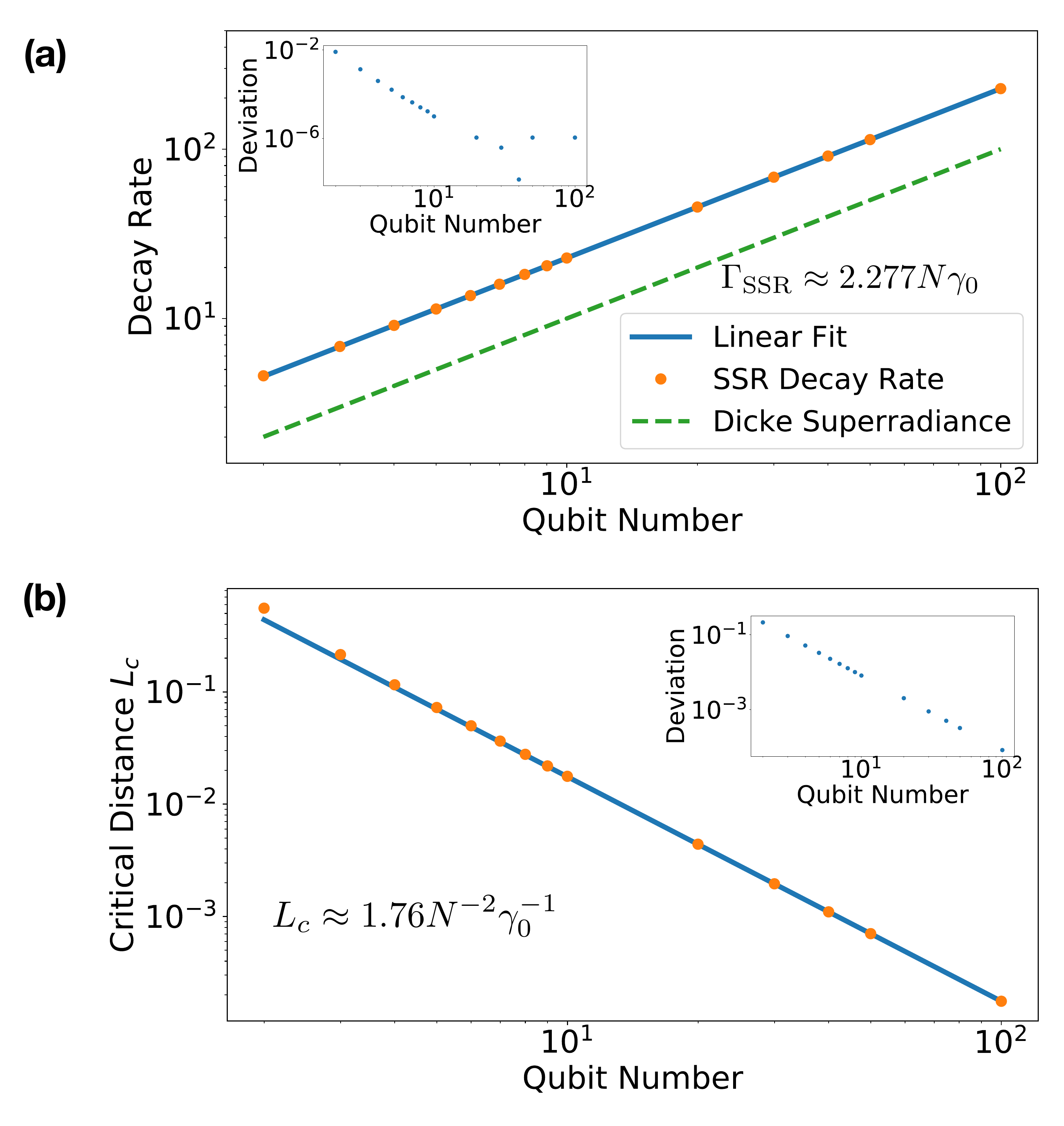}
    \caption{The scaling of (a) the SSR decay rate $\Gamma_{\rm SSR}$ and (b) the critical distance $L_c$ w.r.t. the qubit number $N$. The inset figures show the relative deviation between the fit and the numerical values ($\rm \left|\frac{fit-data}{data}\right|$). The uncertainty in the scaling of the SSR decay rate is well within the errors of the numerical maximization algorithm, whereas for the critical distance there is a significant deviation from the fit even in large $N$. We note that, unlike Dicke superradiance, the SSR does not scale exactly linearly for small $N$. There is a significant difference for $N\leq 10$ between the fit and the numerical data.}
    \label{fig:fig3}
\end{figure}

We now ask whether SSR is a general phenomenon that persists beyond the known cases of $N=2$ and $N=3$. To answer this, we repeat the above analysis for various $N$ up to $N=100$. We find that indeed SSR persists for large $N$ and that $\Gamma_{\rm SSR}$ scales linearly, as shown in  Fig. \ref{fig:fig3} (a).
The scaling is similar to that of Dicke superradiance. Fig. \ref{fig:fig3} (b) shows how the critical distance $L_c$ scales with $N$. The fit is almost linear, but deviates for $N<10$.

To study the closeness of the linear fits, we plot the deviation of the numerical results in Fig. \ref{fig:fig3}. For $\Gamma_{\rm SSR}$, the deviation hovers around $\sim 10^{-6}$ for large $N$, which is within the numerical precision of our algorithm. For  $L_c$, the fit is not as close: the $N^{-2}$ scaling fits the data within $~0.1\%$. For more  evidence for this scaling, we  turn to analytical investigations.

\begin{figure}
    \centering
    \includegraphics[width=\columnwidth]{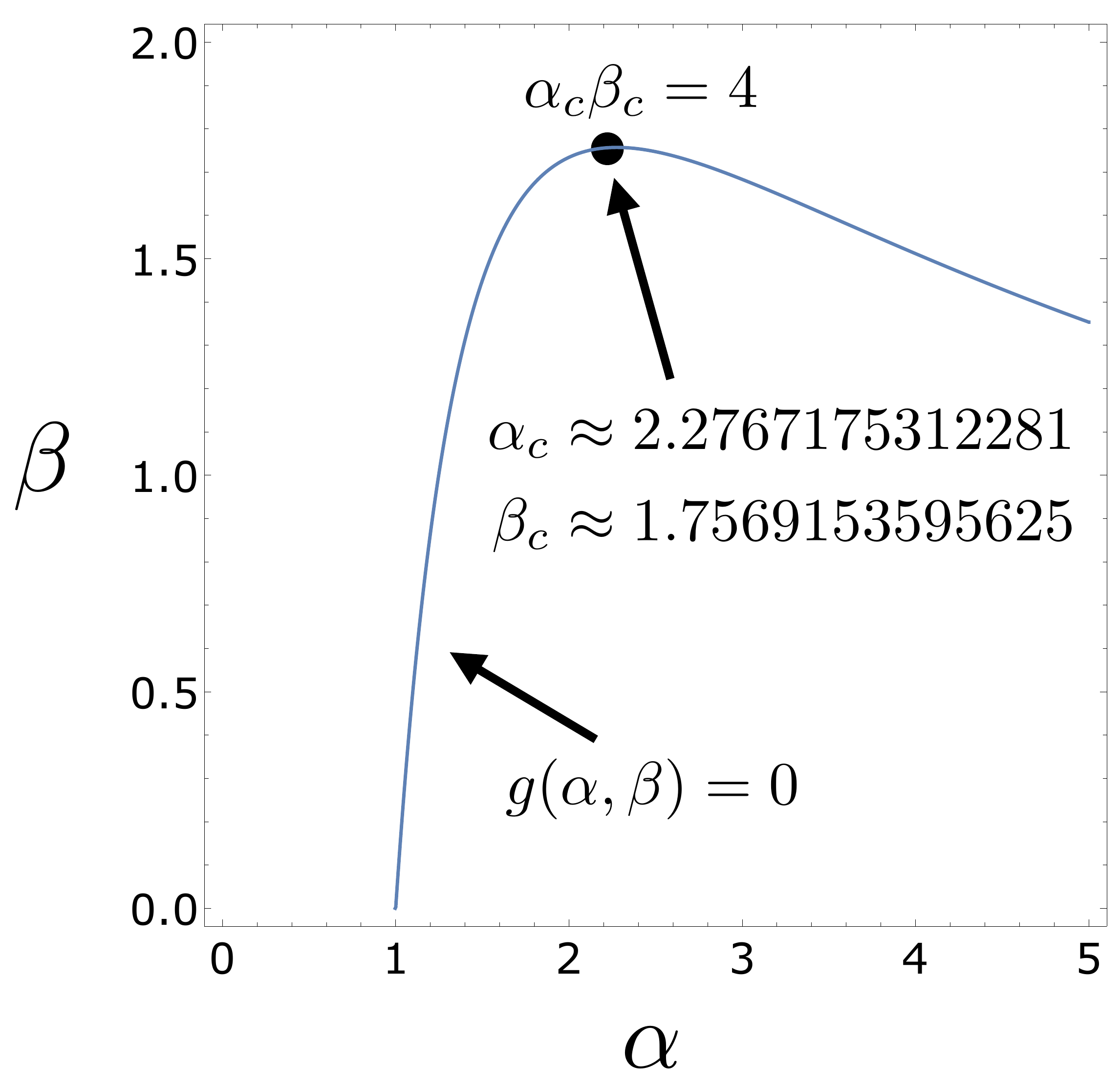}
    \caption{The contour plot of $g(\alpha,\beta)=0$. Here, the assumption $\Delta_k = -i 0.5\alpha N \gamma_0$ and $L_c= \beta N^{-2} \gamma_0^{-1}$ is valid as long as we pick a pair $(\alpha,\beta)$ on the contour. For $\beta > \beta_c$, there is no such pair. For $\beta < \beta_c$, there are two such pairs corresponding to two distinct poles (one exclusively non-Markovian, one Markovian-like). The pole with smaller $\alpha$ corresponds to $\Gamma_u$, but is not the SSR decay rate since larger $\alpha$ can be achieved for $\Gamma_u=\alpha N \gamma_0$ (and $\Gamma_{\rm SSR}= \max_\beta \text{Re}[\Gamma_u]$). For $\beta=\beta_c$, there is only one such pair, corresponding to $\Gamma_{\rm SSR}$ and $L_c$. Note that for the pair $(\beta_c,\alpha_c)$, the equation $\alpha_c \beta_c=4$ is an exact one.}
    \label{fig:fig4}
\end{figure}

A closed form solution exists for the transmission and reflection coefficients related by the transfer matrix as in (\ref{eq:transfermatrix}) \cite{tsoi2008quantum}. Applying the superradiant condition $\Omega L = \pi n$, the characteristic set of equations corresponding to the poles becomes

\begin{subequations} \label{eq:analytic}
\begin{align}
    \cos(\lambda) &= \cos(p L) +\frac{\gamma_0}{2p} \sin(p L), \\
    (p+i\gamma_0/2)\sin(\lambda N) &=e^{ip L} p \sin(\lambda (N-1)).
\end{align}
\end{subequations}
Here, $\lambda$ is a complex number that relates both equations and $p$ relates to the collective decay rate as  $\Gamma = 2i p$. Now, let us set $p=-i\frac{\alpha}{2} N \gamma_0$ ($\Gamma= \alpha N \gamma_0$) and $L=\beta N^{-2} \gamma_0^{-1}$ from the numerical fit we obtained in Fig. \ref{fig:fig3}. Here, the pair $(\alpha,\beta)$ are free parameters to be determined. Then, neglecting terms of order $O(N^{-1})$, the set of equations in (\ref{eq:analytic}) becomes 
\begin{equation}
\begin{split}
    g(\alpha,\beta)= 2\alpha \tau \cosh(\tau) - (2+\alpha^2 \beta ) \sinh(\tau)=0,
\end{split}
\end{equation}
with $\tau =0.5\sqrt{\beta(4+\alpha^2 \beta)}$. We note that the coefficient of $O(N^{-1})$ term is nonzero, which explains the deviations for $N<10$. The SSR decay rate scales asymptotically for large $N$ in contrast to the exact scaling of the Dicke-superradiance. There are infinitely many $(\alpha,\beta)$ pairs that satisfy this equation, which has been illustrated in Fig. \ref{fig:fig4}. The most important observation is that there exists a $\beta_c$ such that there is no solution \footnote{This has been checked for a large interval of $\alpha$ and $\beta$ values numerically.} to $g(\alpha,\beta)=0$ for $\beta>\beta_c$ and there are two pairs $(\alpha_{s/l},\beta)$ (s/l stand for small/large) for $\beta<\beta_c$. In this case, the Markovian-like decay rate scales as $\Gamma_u = \alpha_s N \gamma_0$ and the $N+1$th (exclusively non-Markovian) decay rate scales as $\Gamma_{\rm NM}=\alpha_l N \gamma_0$. As $\beta \to \beta_c$, both $\Gamma_u$ and $\Gamma_{\rm NM}$ come closer and right at $\beta=\beta_c$, $\Gamma_u$ is maximized and has the same real part as $\Gamma_{\rm NM}$. Consequently, $\Gamma_{\rm SSR}=\max_{L} \text{Re}[\Gamma_u]$ is obtained at the position $\frac{\partial g(\alpha,\beta)}{\partial \alpha}=0$. This corresponds to the condition $\alpha_c \beta_c=4$, with $c$ standing for the critical values. The characteristic equation can be solved for this condition and the resulting $(\alpha_c,\beta_c)\approx (2.277,1.76)$ pair agrees with the numerical fit in Fig. \ref{fig:fig3}, confirming our numerical analysis. It is important to note that the crucial step here is the Ansatz that $p \sim N \gamma_0$ and $L \sim N^{-2} \gamma_0^{-1}$, which was obtained from the numerical analysis in the first place.

This analytical investigation also explains  another interesting phenomenon. The imaginary part of $\Gamma_{\rm SSR}$ {is responsible for} the collective energy level shifts. For $L \ll L_c$, the superradiant decay rate $\Gamma_u$ is strictly real and does not have an imaginary component, whereas the $N+1$th exclusively NM collective decay rate is negligibly far away from the origin. This justifies the $N$-pole approximation performed in \cite{zheng2013persistent}, where the $N+1$th pole is so far away from the first $N$ such that it can be neglected. However, for $L> L_c$, the $N$th and $N+1$th decay rates become degenerate. They have the same real parts, but also imaginary parts with the opposite sign. The fact that the decay rates acquire imaginary components for $L>L_c$ suggests a shift in collective energy levels analogous to the Lamb shift. A similar observation has been made in \cite{zheng2013persistent} for $N=2$ qubits. Furthermore, we point to another interesting phenomenon. The energy levels corresponding to the $N$th closest pole to the origin changes discontinuously around the superradiant condition $\Omega L = n \pi$ with changing $L$ for $L>L_c$. This has been plotted in Fig. 4 of \cite{zheng2013persistent} for $N=2$, where increasing the distance between qubits can lead to discontinues changes in the energy level shifts around $L$ values where $\Omega L = n \pi$. We observe here that this is a more general phenomenon for any $N$. We plan to investigate  the shifts in the collective energy levels due to non-Markovian effects further in future work. 

In summary, the SSR decay rate scales linearly ($\Gamma_{\rm SSR} \sim N\gamma_0$) and the critical qubit separation scales by the inverse square ($L_c \sim N^{-2}\gamma_0^{-1}$). While the time-delayed quantum coherent feedback leads to a stronger decay of the collective system, the scaling of the decay rate is still linear with increasing $N$. Furthermore, while $\Gamma_{\rm Dicke} \sim N\gamma_0$ fits exactly for any $N$ in the case of the Dicke superradiance, the value $\Gamma_{\rm SSR} \sim 2.277 N\gamma_0$ fits well  asymptotically for large $N$.  We don't yet know the origin of the linear scaling but we hope to investigate this interesting phenomenon further. As a starting point, it is important to realize that the overall size of the system is $\approx NL_c \sim 1.76N^{-1}\gamma_0^{-1}\ll \gamma_0^{-1}$. This means that light initially emitted by a qubit at one end of the chain travels all the way to the other end and back in less time than the half-time of that individual qubit. Consequently, even photon-mediated interactions between the first and $N$th qubits are in fact significant in the emergence of the SSR phenomenon. Conversely, when qubits are separated by large distances, the collective decay rates become subradiant in agreement with the findings of \cite{zheng2013persistent}. Hence, SSR is only observed when the qubits are moderately separated ($L \approx L_c$), where both collective interactions and time-retardation effects are at play.

When it comes to applications, the non-Markovian system provides additional control parameters. In Markovian 1D waveguide QED, the decay rates depend on the coupling and the phase parameter $\Omega L$. In the non-Markovian regime, however, the decay rates also depend on the qubit separation $L$ and frequency $\Omega$ independently. Furthermore, the distance can enhance the decay rate beyond Dicke superradiance. By tuning the qubit separation, one can tune the interaction between photons and multi-qubit systems, which might  have application for quantum technologies such as quantum memories \cite{kimble2008quantum,lvovsky2009optical,saglamyurek2018coherent}, quantum gates \cite{combes2018two} and pulse shaping \cite{dinc2019exact}. Furthermore, our formalism makes it possible to study systems with many qubits. We thus expect our approach to be useful for  designing new quantum technologies. 

Research at the Perimeter Institute is supported by the Government of Canada through the Department of Innovation, Science and Economic Development Canada, and by the Province of Ontario through the Ministry of Research and Innovation. We acknowledge the support of the Natural Sciences and Engineering Research Council of Canada (funding reference number RGPIN-2016-04135).

\end{document}